# AN ANALYSIS OF IRAS IDENTIFIED HII REGIONS AND THEIR RADIO PROPERTIES

V.A. Hughes

*Astronomy Group, Department of Physics, Queen's University, Kingston, Ontario K7L 3N6, Canada.*

and

G.C. MacLeod

*Hartebeesthoek Radio Astronomy Observatory, P.O. Box 443, Krugersdorp 1740, South Africa.*

## ABSTRACT

To try and confirm the types of object in the list of 2298 potential HII regions identified by Hughes & MacLeod from the IRAS Point Source Catalog, we selected a sample of 82 for observing at the VLA. We selected half with values of $Y = \log(F_{25}/F_{12}) \geq 0.8$, and for control purposes, half with values of $0.3 \leq \log(F_{25}/F_{12}) \leq 0.5$. 78 radio sources were detected, and of all the objects, 72% had at least one associated radio source. Most of the radio sources had diameters of $< 3''$, which was the limit to the angular resolving power of the survey. Those with larger values of Y had significantly larger values of peak radio, integrated radio, and 100$\mu$m flux densities than those with smaller Y. Also, they generally had associated masers, and thus were most likely young compact HII regions containing star forming regions. Those with smaller Y tended not to have associated maser activity and are probably older HII regions, or stars with high IR and ionizing radiation, such as T-Tauri type stars. However, all 2298 objects appear to have a strong galactic concentration. Some comparison is made with the selection of 1717 objects identified by Wood & Churchwell as HII regions, but their selection criteria were somewhat different, and they actually form a subset of ours.

*Subject headings:* IRAS Sources - Radio HII regions

## 1. Introduction

The IRAS Point Source Catalog (IPSC) is a listing which contains flux densities at 12, 25, 60, and 100$\mu$m, of over 250,000 point sources, and provides an unprecedented resource for the unbiased identification and study of astronomical objects. For example, late type stars have been identified by van der Veen & Habing (1988), planetary nebulae by Jourdain et al. (1990). However, the catalog is particularly useful in studies of star formation, since the latter takes place at the centre of molecular clouds which have high optical obscuration. Since only infrared and radio waves can penetrate the clouds the use of the IPSC together with radio observations, are two of the ways by which the regions can be studied. The chief difficulty is how to interpret the data.

Some initial identifications of optically obscured objects were carried out by Beichman et al. (1986) who selected 95 molecular cloud cores from the lists of Myers, Linke & Benson (1983), Myers and Benson (1983), and Benson (1983). The position of the cores were searched for IRAS sources, and 47 were found. From the selection criteria it appeared that the IR sources were associated with newly formed or forming low-mass stars, and about one-third of the sources were associated with visible stars that had properties that resembled T-Tauri stars. Of the remainder, some also could have been associated with T-Tauri stars, but appeared to be embedded in the clouds. Parker (1991), and Parker, Padman & Scott (1991) also correlated IRAS sources with the dark molecular clouds cataloged by Lynds (1962). They observed that a remarkably large 70% of the clouds had molecular outflows, but concluded again that the clouds contained T-Tauri stars with mass typically $\sim 1 M_\odot$. It was pointed out that the IRAS data alone are not sufficient to distinguish between the youngest "protostellar" sources and certain more evolved pre-main sequence stars.

Wood & Churchwell (1989a, hereafter WC1) observed known HII regions using the Very Large Array of the the National Radio Astronomy Observatory [1]. They identified seventy-five ultra-compact (UC) HII regions, which for the resolution of $0\rlap{.}''4$, had diameters of $\lesssim 0.1$ pc, and displayed five different morphologies. They showed that these regions, which they described as massive star forming regions, could in general be differentiated from other objects in the IPSC by using the color criteria of log $(F_{60}/F_{12}) \geq$ 1.3 and log $(F_{25}/F_{12}) \geq 0.57$, where $F_\lambda$ is the flux density at wavelength $\lambda$ ($\mu$m). Wood & Churchwell (1989b, hereafter WC2) then used these criteria to search the whole of the IPSC, rejecting those which had only upper limits at either 60 $\mu$m or 25 $\mu$m, and listed 1717 objects, which they identified chiefly as OB-type stars There was some confirmation of the type of object, since the distribution in latitude falls off quite rapidly, with a scale height of $0\rlap{.}°6 \pm 0\rlap{.}°05$.

Hughes & MacLeod (1989, hereafter HM) used a somewhat different method to select HII regions. They used the IRAS sources associated with known HII regions, planetary nebulae and reflection nebulae and determined their colors. With a view to selecting all HII regions, rather than those associated with the more massive stars, they used principally values of Y = $\log(F_{25}/F_{12}) \geq 0.0$, and X = $\log(F_{60}/F_{25}) \geq 0.0$. A list of 2298 possible HII regions was identified to within a confidence factor of > 70%, and again, there was an indication that they were a galactic class because of their close grouping to within about $\pm$ 3° of the galactic plane. It was a closer grouping than that of the 997 suspected planetary nebulae selected by their other criteria, which had about twice the spread, a trend that might be expected. However, when the possible HII regions were used in a latitude-longitude plot, they appeared to show more details in galactic structure than the objects identified by WC2, in particular there was an increase in number of objects, going up to higher latitude at $\ell$ = 80° which was not obvious in WC2.

One method of identifying an HII region is by way of measuring its radio properties. Accordingly, a representative number of 41 objects with large Y were selected from HM for observing with the VLA, and as a control, an equal number with small Y. This paper describes the selection criteria, the results of the observations, and attempts to use these to define the types of object. In §2, the selection criteria used by HM is described, and §3 describes the results of the observations of 82 of these using the VLA. §4 describes the selection criteria of WC2, a discussion follows in §5, and the conclusions are summarized in §6.

---

[1] The National Radio Astronomy Observatiory is operated by Associated Universities Inc., under co-operative agreement with the National Science Foundation.

## 2. The Selection Criteria.

HM first selected known objects, such as HII regions, reflection nebulae, and planetary nebulae, and plotted the related IRAS sources on a color-color diagram of Y = log ($F_{25}/F_{12}$) vs. X = log ($F_{60}/F_{25}$). This identified areas on the diagram which contained the different objects. The criteria then chosen for HII regions were Y $\geq$ 0.0, X $\geq$ 0.0, and -10° $\leq$ b $\leq$ +10°. There were also conditions set by flux-quality numbers, namely $F_{100} \geq$ 50 Jy for FQ = 2 or 3, and $F_{100} \geq$ 150 Jy for FQ = 1. With the condition that no more than one confusion flag per source be turned on, a list of 2298 sources was obtained. Identifying those that could be seen optically, and taking into account possible contamination by supernova remnants, it was estimated that of this number, 77% were HII regions, 5% were planetary nebulae, 7% were stellar sources, 2% were reflection nebulae, 1% were extragalactic, and 9% were supernova remnants. It is possible that some in the list could actually be solar-type star-forming regions, based on the search by Beichmann et al. (1986), who identified a number of objects as T-Tauri-type stars, or stars having infrared properties similar to T-Tauri stars, but as discussed in HM, this was estimated at < 1% of the total.

A plot of latitude against longitude clearly showed that the objects were galactic, with most lying inside the range of $\pm 3°$. Of interest, there does appear to be some variation about the Galactic plane, and possible features, such as an apparent Cygnus arm at $\ell \sim 80°$ where a number of objects are seen at latitudes up to +6°. Also of interest is the fact that 997 planetary nebulae were identified, but when plotted against longitude, they show the much greater latitude spread of $\sim \pm 10°$, and indicate a possibly different population.

## 3. Observations.

It was quite clear that to observe all 2298 objects at the VLA was not feasible. Accordingly, a representative number was selected, based on choosing 41 with large Y, and as a control, an equal number with small Y. It was considered that the total of 82 could be observed inside a reasonable time of 12 hours. For those with large Y, objects had to be north of declination -36°, they had to have good flux quality factors at all bands, they were the strongest 100 $\mu$m sources, and a check was made that they were not known to have been observed previously at the VLA. The criterion that Y $\geq$ 0.8 then gave the required number of sources of 41. Similar criteria were used for low values of Y, which determined limits of 0.3 $\leq$ Y $\leq$ 0.5. Observations were made on these 82 sources at 1989 July 29, using the "C" configuration at the wavelength of 6 cm, angular resolution 3″. Each object was observed for 7.5 minutes, giving a typical rms noise level of $\sim$ 0.2 mJy per beam. Secondary calibrators were observed at intervals of 30 min, and flux densities related to 3C286. Standard reduction techniques were used, involving an editing session, and MX. The program IMFIT was then used to determine position, peak flux density and total flux density.

The results are given in Table 1 for Y $\geq$ 0.80, and Table 2 for 0.3 $\leq$ Y $\leq$ 0.50. Each table is divided into three parts. In part (a) is listed the sources that were clearly detected as small diameter objects, part (b) lists those whose radio positions were > 3′ from the IRAS positions, and part (c) lists those which were rejected, either because they were below the detection level, or were too diffuse and extended to give a significant flux density. For both Tables, where there was more than one radio source associated with an IRAS source, entries have been made for each. Columns 1-6 give the IRAS data, consisting of name, RA, dec, values for Y = log($F_{25}/F_{12}$), values for X = log($F_{60}/F_{25}$), and 100 $\mu$m flux density. Columns 7-11 list the radio results consisting of RA, dec, peak flux density, total flux density, and offset of the radio from the IRAS position. Column 12 contains comments. In cases where maser emission has been detected from $CH_3OH$ at 6.6 GHz, OH at 1.6 or 4.7 GHz, or $H_2O$ at 22 GHz, the references to the detection are given in columns 13-16. If the reference is written inside square brackets, it indicates that observations were made, but no line was detected. From the list of objects clearly resolved, average values of radio data and of 100 $\mu$m flux density are given for the IRAS sources, and for the total number of radio sources.

The chief point of interest is that in Table 1(a), 24 IRAS sources, or 59% of the total, had one or more associated radio source, although if we include those where the offset was greater than 3 arcmin, the number of associations is 27, or 66%. In Table 2 the number of close associations was 31, or 76%. Thus, of the 82 IRAS sources, 55, or 67% showed close association with a radio source, and if the total possible number of 59 is used, the percentage of associations rises to 72%. The fact that for the majority of these objects, the total radio flux density was approximately equal to the peak flux density, shows that their size was not much greater than about $3''$. Comparing the unresolved sources in Tables 1 and 2, those for $Y \geq 0.8$ showed average values of peak radio, average radio, and 100 $\mu$m flux densities which were, respectively, factors of approximately 10, 3, and 3 greater than the equivalent values for $0.3 \leq Y \leq 0.50$.

The success rate in the radio detection of objects with large Y was as might be expected from the selection criteria used in HM for HII regions, and the presence of maser activity associated with the selected objects can be taken to confirm that they were HII regions; those with low Y did not in general show maser activity, yet approximately the same percentage were radio sources. However, a comparison of the galactic distribution of the objects in HM with those selected by WC2 showed differences in detail, yet in both cases the objects appeared to be components of the galaxy. This suggested that a comparison be made with the selection criteria used by WC2, with a view to determining the nature of the objects in each list.

## 4. The Selection Criteria of WC2.

The procedure used by WC2 was first to use ultra-compact (UC) HII regions identified by WC1. These sources had been detected in the microwave Galactic plane survey of Wink, Altenhoff & Mezger (1982), in the millimeter-wavelength observations by Wood, Churchwell & Slater (1988), and in Wood et al. (1988). The criteria were (1) the presence of a small radio continuum source, (2) a radio spectrum consistent with free-free emission, and (3) strong FIR emission with a spectral shape similar to that seen in the UC HII regions that were known at that time. Observations with the VLA showed radio detection in 53 cases, some having multiple radio sources, and null results at 6 cm were obtained in 31 cases. Based on the fact that the IRAS sources had strong FIR emission, and the similarity of the FIR shapes, it was concluded that the radio objects were HII regions ionized chiefly by O-type stars.

WC2 then plotted these well defined UC HII regions on color-color diagrams of $\log(F_{100}/F_{60})$ vs. $\log(F_{25}/F_{12})$ and $\log(F_{60}/F_{12})$ vs. $\log(F_{25}/F_{12})$. They then chose all IPSC objects in a $2° \times 2°$ box centered in the Galactic plane at longitude $40°$, and all sources in a strip of sky between $13^h00^m$ and $13^h10^m$ in right ascension and plotted these also on the same diagram. It was clear that those objects which were identified as UC HII regions formed a particular region of the diagram, and in fact they clustered around $\log (F_{100}/F_{60}) = 0.3$, $\log (F_{60}/F_{12}) = 2.0$, and $\log (F_{25}/F_{12}) = 1.0$. From the distribution of their flux densities, the median value at 100 $\mu$m was 330 Jy, in comparison with a median value of 5,000 Jy for the known UC HII regions. It was concluded that they detected all B0 ZAMS stars and hotter at 60 and 100 $\mu$m, all B0 ZAMS stars and hotter at 25 $\mu$m, and all O9.5 ZAMS stars and hotter at 12 $\mu$m, anywhere in the Galaxy. Thus, their sample will be dominated by high-luminosity O-type stars, but will also contain some early B-type stars.

Since the plot of $\log(F_{60}/F_{12})$ showed the largest spread in the diagrams, the color criteria chosen by WC2 to identify embedded massive stars were $\log (F_{60}/F_{12}) \geq 1.30$ and $\log (F_{25}/F_{12}) \geq 0.57$. The result was the identification of 1717 candidates. When these objects were arranged on a latitude-longitude plot, the objects appeared clearly to be a Galactic plane population, centered at $b = 0°$ to within a single bin of $0°.2$, and falling off exponentially with an angular scale height of $0°.6 \pm 0°.05$.

Thus, the selection criteria for HII regions in the two cases were somewhat different. WC2 used UC HII regions which resulted in a selection of objects with larger values of Y. In fact, as we shall see, they selected objects which coincided with the larger values of Y used in this paper. The selection made by HM used a lower limit to Y, which selected flatter spectra and thus lower flux densities at each band. This reduced the effect of distance but possibly included more objects of differing types. However, the percentage of radio detections in this paper was essentially the same, regardless of the value of Y, so the question is asked "What are the objects being selected in each case?". These matters will be further discussed in § 5.

## 5. Discussion.

The first point of interest is that for a representative number of IRAS sources selected from HM as containing HII regions, between 67% and 72% have radio sources associated with them, the majority having small angular size. None of our objects was in the list by Parker, Padman & Scott (1991), as expected because of the selection criteria, and there was no correlation with the results of Beichman et al. (1986) for the reasons stated previously.

To compare the selection criteria by WC2 with that by HM, we have plotted our representative objects on a diagram of Y vs X in Figure 1, and in Figure 2 the same objects in a diagram Y vs X', where we here define $X' = \log(F_{60}/F_{12})$ the colors used by WC2. We note that the objects chosen from HM with $Y \geq 0.8$, also have values of $X' \geq 1.3$, $Y \geq 0.57$, the colors used by WC2, but the percentage of radio detections of the HM sources is not very dependent on whether or not $Y \geq 0.8$, or $0.3 \leq Y \leq 0.5$. On the other hand, mean values of the peak, average flux densities, and 100$\mu$m flux densities are appreciably less for $0.3 \leq Y \leq 0.5$. Also, for $Y \geq 0.8$, as shown in Table 1, most sources which are clearly resolved have associated maser activity, and Menten (1991) has shown that 6.6 GHz $CH_3OH$ masers are Class II masers and invariably associated with compact HII regions. The presence of $H_2O$ masers is normally also considered empirical evidence for the presence of young HII regions. For the case where $0.3 \leq Y \leq 0.5$, there is a general failure to detect an $H_2O$ maser line, although no observations of these sources for $CH_3OH$ and OH have been published.

Habing & Israel (1979) have noted that maser activity is associated with young and compact HII regions, but when they expand to a radius greater than 20,000 au, the OH maser condition disappears. Objects with smaller values of Y are not generally accompanied by maser activity, so that they could be larger size HII regions, or they could contain stars with comparatively large IR and ionization fluxes, of which T-Tauri stars are an example. In this respect we refer also to Cepheus A (Hughes & MacLeod 1993) which appears to have both a larger IR and larger ionization flux than can be attributed to normal ZAMS stars.

We have tried to see if different Galactic populations are being observed as follows. HM initially plotted values of Y against X for IRAS sources identified as known HII regions, planetary nebulae (PN's), and reflection nebulae (RN's). Different areas of the color-color diagram contained the different populations, but there was clearly an overlap. As we have mentioned, the final list of 2298 possible HII regions contained also point sources that are associated with other known types. Thus, corruption in the identification of HII regions is expected chiefly from PN's, stellar sources, and SNR's, although PN's have been seen to produce an overall larger spread in latitude at all longitudes. Also, any distribution of stellar objects, unless the objects are young, might be expected to go to higher latitudes; either they will have migrated from the galactic plane or they will be somewhat nearer. It is thus of interest to compare the latitude distribution of the identified objects.

In the case of the 1717 objects identified by WC2, and the 2298 identified by HM, there is a clear concentration towards the galactic plane, indicating that in each case we are dealing with a galactic population. WC2 plotted the distribution of their objects about the plane, and found that they were centered on b = 0° to within a single bin of < 0°.2. An exponential was fitted to the distribution which showed an angular scale height of 0°.6 ± 0°.05. An inspection of their Figure 3 shows no obvious galactic structure, except that there are more objects in the first and fourth quarters than in the second and third.

The 2298 objects identified by HM also show a concentration towards the plane, and a greater density of objects towards the galactic center. As previously mentioned, they are more closely concentrated towards the plane than the 997 planetary nebulae also identified. But in addition, the former do show some evidence of structure, and in particular an increase in numbers up to b = 6° at $\ell = 80°$, in comparison with adjacent longitudes. This we initially suggest could be the Cygnus arm, but it is not apparent in the data by WC2. This is somewhat surprising since the latter should be detecting primarily O-type stars, and these are thought to form in the spiral arms. On the other hand, the smaller values of Y, may define a different population.

The other aspect of the data by HM is that although towards the galactic centre the objects clearly concentrate towards the plane, towards the anticenter they have a much wider distribution in latitude. This is clearly shown in Figures 3, 4, and 5. Figure 3 shows the distribution of those of the 2298 objects which are in the longitude range $0° < \ell < 60°$. They are compared with a gaussian distribution which has a mean at -0°.09, and a standard deviation of $\sigma = 0°.42$, or half-width to half-maximum of 0°.49, but there are wings out to ± 2°.5. Thus they would appear to have a comparatively narrow galactic plane distribution, as expected for HII regions, but with an additional superposed and wider distribution which, as we have previously mentioned, is likely due to SNR's, PN's or nearby objects. The data by WC2 are not sufficiently detailed to show this effect. Figure 4 shows the distribution of those in the range $300° < \ell < 360°$, and again is very similar to that of Figure 3. It is compared with a gaussian having a standard deviation of $0°.5$, and again shows wings out to larger latitudes. Figure 5 shows the distribution over the range of the anticenter, $120° < \ell < 240°$, and although there may be a concentration towards b = 0°, there is generally the broader distribution as is shown when compared with a gaussian having $\sigma = 1°.8$. Thus, although the distribution in the galactic plane in the direction of the center shows objects which are concentrated to $\pm 0°.5$ of the plane, or $\pm 100$ pc for typical distances of 10 kpc, and are presumably distant objects, as could be the case for the objects in the WC2 list, we also appear to be detecting much nearer objects, which if contained within $\pm 100$ pc of the plane, are inside the distance of $\sim 3$ kpc. It is possible that WC2 also have a wider distribution, but it may be masked by their fitting of an exponential curve.

We have tried to define more closely different HM objects from their IR spectra. We used the spectra of HII regions in the optical and radio catalogs by Sharples (1959) and Blitz, Fich, & Stark (1982), those of PN's from the catalogs by Higgs (1971) and Iyengar (1986), and those for reflection nebulae from van den Bergh (1966a, and b). To reduce possible effects due to distance, average values of $\log(F_{25}/F_{12})$, $\log(F_{60}/F_{25})$, and $\log(F_{100}/F_{60})$ were determined. The individual spectra were then normalized so that the maximum flux density of each was $100\mu$m, and the values of the above ratios used to determine mean values of flux density at each wavelength. Table 3 shows the averaged values for the ratios, with standard deviations shown in parentheses, the derived values for flux density, and the average value of flux density at $100\mu$m, (possibly influenced because of the range of distances). The corresponding spectra are plotted in Figure 6. Bearing in mind that these are for observed objects, the spectra for planetary nebulae and reflection nebulae probably represent more closely the true spectra of the objects themselves, since they are principally associated with objects which are not at the center of clouds; the former peak at about $30\mu$m which is the peak for a black body at 100K, while the latter appear to peak at a wavelength $\geq 100\mu$m corresponding to a black body at $\leq 30$K. Also, average values for the $100\mu$m flux density for both are at least an order of magnitude less than for HII regions, though different distances could affect this somewhat. It is to be expected that for HII regions in general the total IR flux density will be greater, since they will be embedded in clouds, and hence their total energy will be spread over a large volume of the cloud, the cloud temperature is expected to be less, and the peak in flux density to move to longer wavelengths.

Thus, when all suspected HII regions are plotted as Y vs X', we expect, as is observed, that the points will extend further upwards and to the right. We have also plotted on the Y vs X curve of Figure 1, the average values determined as above for all three objects, but only the average value for HII regions in Figure 2 since values for the other two are off the diagram to the left. On this basis, it is understandable that most of the objects selected by WC2, and those with larger values of Y in HM do refer to HII regions produced by highly luminous embedded star forming regions, but questionably not all of them. Only about 60% of our objects are radio sources, and we suspect the same percentage for the WC2 objects. But there is also a large number of objects in this survey, principally those with values of $Y \leq 0.5$, which are also detected as radio objects. These could be either expanded HII regions, or RN's though the $100\mu$m flux densities are quite high, or less luminous star forming regions, or T-Tauri stars as suggested by others, or other very young stars which have high luminosity and ionizing radiation.

In this respect, we refer to the work by Beichmann et al. (1986) who identified IRAS objects embedded in molecular cloud cores. They occupy a similar position on the color-color diagram to those of HM, having values of $\log(F_{60}/F_{25}) \sim 0$, and $\log(F_{25}/F_{12}) \sim 0.3$, but generally the flux density at 100 $\mu$m is < 100 Jy. These objects appeared to be associated with T-Tauri type stars. Also, the objects observed by Parker, Padman, & Scott (1991), which were selected from IRAS sources associated with Lynds dark clouds, were generally attributed to stellar type stars of $\sim 1 M_\odot$. On the other hand, 70% of the sources appeared to have outflow activity, which we can attribute to low mass stars, presumably in their protostar stage. We are led to suggest that some of the objects that both WC2 and HM have identified are protostars with excess infrared emission and excess ionization. Such a conclusion was arrived at by Hughes & MacLeod (1993), who suggested that they might be observing later type stars which are in the Hayashi phase.

A more detailed investigation might even show an evolving effect with time, as suggested by MacLeod &

Hughes (1991), but also the lack of detection at radio wavelengths of some sources with $Y \geq 0.8$ could be due to very high densities so that the HII region is very small and compact. Also, a number of theories on the formation of stars more massive than 5 $M_\odot$, suggest that a lower mass star forms, and can reach the main sequence while still continuing to accrete, so that there could be absorption of ionizing photons by gas and dust which is falling onto the star. Thus it may be that there are some objects that are low mass stars of about B0 - B3 which are in the accretion stage and showing the excess infrared radiation as do T-Tauri stars, but have not yet started to show the additional ionization. They are progressing towards B3 stars which are the latest stars that can theoretically be detected by radio emission from their HII regions, and have either not reached this stage, or are progressing towards more massive stars.

One other significant fact is the apparent small size of the radio sources. For the majority which have angular diameters of $\leq 3''$, the maximum size would be 30,000 au or 0.15 pc at the distance of 10 kpc. These would be classified by WC1 as ultra compact (UC) HII regions, but as we have pointed out, we could be dealing with different classes of object.

## 6. Conclusions.

To confirm the identification of the 2298 potential HII regions selected from the IRAS Point Source Catalog, we have observed a sample of 82 with the VLA. The objects chosen were in two equal sets, half had values of $Y \geq 0.8$, and half with $0.3 \leq Y \leq 0.8$. In general, radio objects with large Y had associated OH, $H_2O$ and $CH_3OH$ masers, which are strong indicators that they are HII regions, and they appear to have similar colors to the 1717 objects identified by WC2. Those with small Y in general did not show maser activity, but showed the same percentage of radio detections, although mean values of peak radio, average radio, and 100$\mu$m flux densities were less by factors of 10, 3, and 3 respectively. Although both the 2298 objects selected by HM, and the 1717 objects selected by WC2 had strong galactic concentrations, there are some differences in detail. This suggests that those with the lower values of Y in the HM objects, though still having the same percentage of radio detections, are likely to be objects with comparatively strong IR and radio emission, such as T-Tauri stars, or are possibly similar to Cepheus A (Hughes & MacLeod 1993).

Most of this work has been carried out with the support of an Operating Grant from the Natural Sciences and Engineering Research Council of Canada.

Fig. 1.— Plot of $Y = \log(F_{25}/F_{12})$ vs $X = \log(F_{60}/F_{25})$ for selected IRAS objects. The + signs are for $Y \geq 0.8$, and the × signs for $0.3 \leq Y \leq 0.5$. The open squares represent the averaged values, as described in the text, and the error bars are 1 $\sigma$.

Fig. 2.— As for Figure 1, but for $Y = \log(F_{25}/F_{12})$ vs $X' = (F_{60}/F_{12})$.

Fig. 3.— Distribution of number of IRAS objects, identified as HII regions, in $0°.2$ intervals of latitude, over the longitude range $0° < \ell < 60°$. Asterisks are for IRAS objects, and +'s show a gaussian distribution with $\sigma = 0°.42$.

Fig. 4.— Same as for Figure 3, but for $300° < \ell < 360°$. The gaussian has $\sigma = 0°.5$.

Fig. 5.— Same as for Figure 1, but for $120° < \ell < 240°$. The gaussian has $\sigma = 1°.8$.

Fig. 6.— Representative spectra for three types of Galactic object, determined as described in the text.

TABLE 1
Sources With Y > 0.8

| IRAS Number | R.A. h.m.s. | Dec d.m.s. | Log (25/12) | Log (60/25) | 100μm Jy | R.A. Radio | Dec Radio | Peak mJy | Total mJy | Offset arcmin | Com | CH$_3$OH 6.6 | OH 1.6 | OH 4.7 | H$_2$O 22GHz |
|---|---|---|---|---|---|---|---|---|---|---|---|---|---|---|---|
| (a) SOURCES CLEARLY DETECTED. | | | | | | | | | | | | | | | |
| 0555+163 | 05 55 20 | 16 31 46 | 1.70 | 0.82 | 525 | 55 20 | 31 44 | 1.40 | 1.55 | 0.03 | 1 | [18] | | | 3,4 |
| 2012+410 | 20 12 41 | 41 04 20 | 1.63 | 1.10 | 1947 | 12 41 | 04 17 | 0.90 | 0.85 | 0.05 | 1 | 5 | 7 | 6 | 3 |
| 1909+093 | 19 09 30 | 09 30 42 | 1.39 | 1.13 | 2740 | 09 31 | 30 46 | 30.80 | 171.77 | 0.26 | 3,4 | 11 | 12 | | 10,12 |
| 1850+012 | 18 50 45 | 01 21 09 | 1.28 | 1.45 | 1950 | 50 47 | 21 00 | 8.60 | 8.58 | 0.52 | 1 | 13 | | 3 | |
| 1845-012 | 18 45 40 | -01 29 49 | 1.27 | 1.09 | 3800 | 45 37 | 29 54 | 60.70 | 262.08 | 0.75 | 5 | 11 | 14 | | 10,15 |
| 1805-195 | 18 05 39 | -19 52 34 | 1.13 | 1.54 | 10000 | 05 40.3 | 52 21 | 41.52 | 73.87 | 0.32 | 2 | 11 | 7 | | 10,16 |
| | | | | | | 05 37.5 | 52 41 | 9.15 | 265.37 | 0.48 | 4 | | | | |
| 0223+613 | 02 23 14 | 61 38 46 | 1.12 | 1.24 | 10600 | 23 17 | 38 57 | 459.80 | 735.57 | 0.40 | 6 | 11 | | 6 | 4,16 |
| 2035+412 | 20 35 05 | 41 26 02 | 1.07 | 0.87 | 3240 | 35 04 | 25 53 | 41.80 | 32.68 | 0.24 | 1 | | 7 | | [3] |
| 2022+372 | 20 22 04 | 37 28 2 | 1.06 | 0.87 | 2130 | 22 0 | 28 25 | 100.0 | 446.83 | 0.20 | 7 | [13] | | | [3].17 |
| 1859+041 | 18 59 38 | 04 16 08 | 1.01 | 0.92 | 232 | 59 34.9 | 15 08 | 0.56 | 1.82 | 1.41 | 2 | | | | |
| | | | | | | 59 34 | 15 00 | 1.13 | 3.05 | 1.51 | 2 | | | | |
| 0535+354 | 05 35 06 | 35 49 34 | 0.99 | 1.20 | 413 | 35 06 | 50 40 | 25.60 | 26.74 | 1.10 | 1 | 11 | | | |
| 0610+152 | 06 10 29 | 15 24 49 | 0.98 | 0.87 | 441 | 10 29 | 24 50 | 4.40 | 6.28 | 0.02 | 1 | [13] | | | [3] |
| 1837-0541 | 18 37 16 | -05 41 37 | 0.95 | 1.26 | 1250 | 37 16 | 41 34 | 61.10 | 85.79 | 0.05 | 1 | [13] | | | [3] |
| 1815-164 | 18 16 00 | -16 48 55 | 0.94 | 1.25 | 2850 | 16 03 | 48 00 | 1.70 | -0.29 | 1.16 | 4 | | | | 3,16,17 |
| 1938+235 | 19 38 53 | 23 57 36 | 0.93 | 1.37 | 432 | 38 53 | 57 40 | 1.60 | 2.21 | 0.07 | 1 | 13 | | | |
| 1815-155 | 18 15 55 | -15 50 10 | 0.90 | 0.88 | 712 | 15 54.6 | 50 08 | 1.91 | 2.84 | 0.07 | 2 | [3],[4] | | | |
| | | | | | | 15 56 | 50 23 | 2.47 | 2.97 | 0.42 | 1 | | | | |
| | | | | | | 15 54.7 | 50 40 | 4.26 | 5.09 | 0.51 | 1 | | | | |
| 1623-241 | 16 23 32 | -24 16 56 | 0.88 | 0.92 | 4650 | 23 33 | 16 45 | 10.40 | 12.72 | 0.29 | 8 | | | | |
| 1808-173 | 18 08 56 | -17 32 14 | 0.87 | 1.44 | 3150 | | | 1.40 | 70.49 | | 9 | 11 | 7 | 6 | 3,16 |
| 1840-034 | 18 40 53 | -03 48 01 | 0.85 | 0.93 | 866 | 40 53 | 50 51 | 52.30 | 164.78 | 2.83 | 2 | | | | [3] |
| 1751-260 | 17 51 46 | -26 09 20 | 0.84 | 1.04 | 1060 | 51 48.4 | 09 08 | 4.12 | 9.87 | 0.58 | 2 | | | | |
| | | | | | | 51 49 | 09 17 | 76.57 | 100.99 | 0.68 | 2 | | | | |
| | | | | | | 51 49.4 | 09 30 | 3.26 | 7.29 | 0.78 | 2 | | | | |
| 1835-064 | 18 35 50 | -06 47 11 | 0.84 | 1.20 | 987 | 35 48 | 47 14 | 5.28 | 8.16 | 0.50 | 1 | [13] | | | [3] |
| 1844-022 | 18 44 47 | -02 29 15 | 0.81 | 1.10 | 496 | | | 61.90 | 84.44 | | 1 | [13] | | | 16 |
| 1921+145 | 19 21 29 | 14 58 57 | 0.80 | 1.04 | 379 | 21 30 | 59 00 | 1.00 | 238.15 | 0.25 | 10 | | | | [3] |
| 1853+004 | 18 53 18 | 00 47 26 | 0.80 | 0.98 | 415 | 53 18 | 47 30 | 9.30 | 15.79 | <0.08 | 11 | | | | [3] |
| | | | | | | | | | | | | | | | |
| Sum | | | | | 55265 | | | 1085.20 | 2848.32 | 15.48 | | | | | |
| No. IRAS Sources | | | | | 24 | | | 24 | 24 | 24 | | | | | |
| Avg. per IRAS Source | | | | | 2303 | | | 45.22 | 118.68 | 0.65 | | | | | |
| Tot. No. of Sources | | | | | | | | 30 | 30 | 30 | | | | | |
| Avg. per Source | | | | | | | | 36.17 | 94.94 | 0.52 | | | | | |
| | | | | | | | | | | | | | | | |
| (b) SOURCES WITH OFFSET > 3′. | | | | | | | | | | | | | | | |
| 0548+254 | 05 48 05 | 25 45 29 | 2.07 | 1.11 | 645 | 48 14 | 48 26 | 30.20 | 36.78 | 3.58 | 1 | 1 | | | 2,[3],[4] |
| 0537-073 | 05 37 31 | -07 31 59 | 1.55 | 1.25 | 269 | 37 39.6 | 29 23 | 1.18 | 3.55 | 3.27 | 2 | | | | [8], 9, 10 |
| | | | | | | 37 40.3 | 28 50 | 11.14 | 12.45 | 3.86 | 1 | | | | |
| | | | | | | 37 40.7 | 28 34 | 3.05 | 6.62 | 4.08 | 2 | | | | |
| 1815-155 | 18 15 55 | -15 50 10 | 0.90 | 0.88 | 712 | 15 41.6 | 47 20 | 4.39 | 6.31 | 4.29 | 2 | | | | |
| 0049+561 | 00 49 28 | 56 17 28 | 0.87 | 1.39 | 1166 | 49 47 | 14 32 | 14.10 | 22.67 | 3.94 | 1 | | | | 4,8,16 |
| 1835-064 | 18 35 50 | -06 47 11 | 0.84 | 1.20 | 987 | 35 33.5 | 50 35 | 14.97 | 17.58 | 5.32 | 1 | | | | |
| | | | | | | 35 26.6 | 48 38 | 5.12 | 8.49 | 5.99 | 1 | | | | |
| | | | | | | | | | | | | | | | |
| Sum | | | | | 3779 | | | 84.15 | 114.44 | 34.33 | | | | | |
| No. IRAS Sources | | | | | 5 | | | 5 | 5 | 5 | | | | | |
| Avg. per IRAS Source | | | | | 756 | | | 16.83 | 22.89 | 6.87 | | | | | |
| Tot. no. of Sources | | | | | | | | 8 | 8 | 8 | | | | | |
| Avg. per Source | | | | | | | | 10.52 | 14.31 | 4.29 | | | | | |

TABLE 1—*Continued*

| IRAS Number | R.A. h.m.s. | Dec d.m.s. | Log (25/12) | Log (60/25) | 100$\mu$m Jy | R.A. Radio | Dec Radio | Peak mJy | Total mJy | Offset arcmin | Con | CH$_3$OH 6.6 | OH 1.6 | OH 4.7 | H$_2$O 22GHz |
|---|---|---|---|---|---|---|---|---|---|---|---|---|---|---|---|
| (c) SOURCES REJECTED |
| 2219+633 | 22 19 51 | 63 36 33 | 1.98 | 0.87 | 400 | | | 0.00 | 0.00 | | 12 | | | | 3,8 |
| 0533-062 | 05 33 53 | -06 24 02 | 1.53 | 1.10 | 488 | ORION | | 0.70 | | | 4 | | | | [8] |
| 2006+355 | 20 06 17 | 35 50 32 | 1.42 | 0.84 | 343 | | | | 0.30 | | 12 | | | | 3 |
| 1831-084 | 18 31 46 | -08 45 47 | 1.20 | 0.99 | 1450 | W41 | | 2.10 | | | 4 | 13 | | | [3] |
| 2022+415 | 20 22 46 | 41 54 29 | 1.09 | 0.84 | 459 | | | 0.60 | | | 4 | | | | [3] |
| 1831-082 | 18 31 04 | -08 25 56 | 1.06 | 1.03 | 571 | | | | | | 12 | 13 | | | [3] |
| 0033+631 | 00 33 53 | 63 12 32 | 1.07 | 1.23 | 683 | | | 0.30 | | | 12 | 18 | | | 4,[8],16 |
| 1921+142 | 19 21 22 | 14 24 58 | 1.01 | | 26760 | W51 | | 1952.00 | | | 4 | 11 | 14 | 6 | 15,16 |
| 1752-243 | 17 52 45 | -24 39 55 | 0.97 | 1.02 | 809 | | | 5.30 | | | 4 | | | | 3 |
| 1941+233 | 19 41 04 | 23 36 54 | 0.88 | 0.96 | 1630 | 41 06 | 35 50 | 0.70 | | | 12 | 11 | | | 10,16 |
| 1853+074 | 18 53 46 | 07 49 30 | 0.88 | 0.99 | 4230 | 54 40 | 53 40 | 43.00 | | 4.30 | 4 | | | | 16 |
| 1813-194 | 18 13 25 | -19 42 25 | 0.86 | 1.01 | 1550 | | | | | | 12 | 13 | | | 3 |
| 2226+624 | 22 26 47 | 62 44 22 | 0.80 | 1.07 | 275 | | | 0.40 | | | 12 | | | | [3],[8] |
| 1835-053 | 18 36 00 | -05 37 49 | 0.80 | 1.41 | 1690 | 36 03.4 | 34 20 | 1.80 | | 3.58 | 4 | | | | |
| | | Average | 1.11 | 1.03 | 2953 | | | | | | | | | | |

COMMENTS
| | | |
|---|---|---|
| 1 Unresolved | 2 Partly resolved | 3 Structure |
| 4 Diffuse emission | 5 Cometary shape | 6 Elongated |
| 7 SNR?, Diam 20" | 7 Slight elongation | 9 Extended, 2' |
| 10 Extended, 1.2x1.5arcmin | 11 Binary, sep. 7" | 12 Not Detected |

TABLE 2
Sources with $0.5 > Y > 0.3$

| IRAS Number | R.A. h.m.s. | Dec d.m.s. | Log 25/12 | Log 60/25 | 100μm Jy | R.A. Radio | Dec Radio | Peak mJy | Total mJy | Offset arcmin | Com | CH$_3$OH 6.6 | OH 1.6 | OH 4.7 | H$_2$O 22GHz |
|---|---|---|---|---|---|---|---|---|---|---|---|---|---|---|---|
| (a) SOURCES CLEARLY DETECTED. | | | | | | | | | | | | | | | |
| 2009+364 | 20 09 55 | 36 40 35 | 0.50 | 1.08 | 270 | 09 55 | 40 36 | 4.80 | 36.92 | 0.02 | 2 | | | | |
| 0142+640 | 01 42 05 | 64 01 01 | 0.48 | 1.06 | 233 | 42 06 | 01 00 | 3.10 | 58.68 | 0.11 | 3 | | | | [3] |
| 1804-215 | 18 04 23 | -21 53 27 | 0.48 | 1.40 | 1020 | 04 24 | 53 40 | 2.10 | 84.00 | 0.00 | 4 | | | | [3] |
| 1802-214 | 18 02 38 | -21 45 25 | 0.47 | 1.34 | 1190 | | | 1.54 | 23.03 | 0.00 | 4 | | | | [3] |
| 0630+040 | 06 30 53 | 04 02 27 | 0.46 | 1.20 | 949 | 30 54 | 02 30 | 10.20 | 82.92 | 0.00 | 5 | | | | [3],4 |
| 1839-043 | 18 39 40 | -04 31 35 | 0.46 | 1.22 | 615 | 39 38 | 31 36 | 5.00 | 20.85 | 0.50 | 6 | | | | [3] |
| 1819-133 | 18 19 24 | -13 35 55 | 0.44 | 1.17 | 5560 | 19 25 | 35 55 | 6.00 | 42.13 | 0.24 | 7 | | | | |
| 1831-073 | 18 31 15 | -07 38 34 | 0.43 | 1.18 | 394 | | | 0.90 | 1.96 | | 4 | | | | |
| 0048+653 | 00 48 27 | 65 31 40 | 0.43 | 1.18 | 234 | 48 28 | 31 43 | 0.80 | 8.01 | 0.11 | 8 | | | | [3] |
| 2030+384 | 20 30 01 | 38 47 09 | 0.42 | 1.33 | 110 | | | 3.31 | 5.49 | | 4 | | | | |
| 0400+505 | 04 00 04 | 50 52 35 | 0.42 | 1.12 | 273 | 00 05 | 52 35 | 3.10 | 22.50 | 0.16 | 9 | | | | [3] |
| 0629+042 | 06 29 09 | 04 21 44 | 0.41 | 1.07 | 496 | 29 09 | 21 50 | 3.20 | 10.12 | 0.10 | 10 | | | | 3 |
| 1828-101 | 18 28 23 | -10 11 06 | 0.40 | 1.21 | 359 | 28 24.4 | 10 56 | 13.20 | 28.68 | 0.38 | 11 | | | | |
| | | | | | | 28 22 | 11 06 | 3.23 | 8.24 | 0.25 | 11 | | | | |
| 1754-244 | 17 54 51 | -24 41 31 | 0.40 | 1.16 | 355 | 54 51 | 41 25 | 6.20 | 56.31 | 0.00 | 12 | | | | |
| 2040+455 | 20 40 40 | 45 55 04 | 0.39 | 1.19 | 436 | 40 40 | 55 05 | 1.70 | 49.94 | 0.00 | 13 | | | | [3] |
| 1853+014 | 18 53 43 | 01 45 05 | 0.39 | 1.33 | 252 | 53 43.6 | 45 05 | 2.10 | 6.46 | 0.15 | 11 | | | | [3] |
| 1836-051 | 18 36 15 | -05 17 55 | 0.38 | 1.29 | 397 | 36 12.5 | 18 06 | 0.46 | 0.71 | 0.65 | 11 | | | | |
| | | | | | | 36 11.9 | 18 24 | 0.35 | 0.45 | 0.91 | 11 | | | | |
| | | | | | | 36 20.4 | 16 19 | 0.73 | 1.34 | 2.09 | 14 | | | | |
| | | | | | | 36 22 | 16 17 | 1.71 | 2.03 | 2.39 | 14 | | | | |
| 1857+032 | 18 57 10 | 03 26 16 | 0.38 | 1.22 | 388 | 57 04 | 26 38 | 2.00 | 77.81 | | 4 | | | | [3] |
| 1816-163 | 18 16 24 | -16 31 39 | 0.36 | 1.52 | 2980 | 16 27 | 29 00 | 6.50 | 7.57 | 2.75 | 11 | | | | 16 |
| 0519+335 | 05 19 46 | 33 55 38 | 0.36 | 1.24 | 579 | 19 51 | 55 36 | 3.53 | 7.87 | 1.04 | 14 | | | | [3],[4] |
| | | | | | | 19 51 | 55 26 | 3.98 | 7.98 | 1.06 | 14 | | | | |
| | | | | | | 19 53 | 55 19 | 6.99 | 14.73 | 1.49 | 14 | | | | |
| 1841-031 | 18 41 09 | -03 13 00 | 0.35 | 1.26 | 381 | 41 10 | 13 00 | 3.10 | 6.05 | 0.25 | 11 | | | | |
| 1920+144 | 19 20 33 | 14 47 36 | 0.34 | 1.24 | 521 | 20 32.5 | 47 36 | 1.50 | 13.49 | 0.12 | 15 | | | | |
| 1831-074 | 18 31 48 | -07 41 19 | 0.34 | 1.52 | 534 | 31 50 | 41 45 | 7.90 | 15.44 | 0.66 | 1 | | | | [3] |
| 1813-162 | 18 13 23 | -16 26 29 | 0.34 | 1.02 | 266 | 13 23 | 26 30 | 3.40 | 227.54 | | 4 | | | | |
| 0600+301 | 06 00 10 | 30 14 19 | 0.33 | 1.25 | 274 | 00 11 | 14 20 | 0.90 | 8.70 | | 4 | | | | 16 |
| 1929+163 | 19 29 36 | 16 37 03 | 0.32 | 1.07 | 256 | 29 36 | 36 54 | 0.56 | 1.80 | 0.15 | 1 | | | | [3] |
| | | | | | | 29 34.5 | 37 08 | 0.75 | 0.91 | 0.37 | 11 | | | | |
| | | | | | | 29 39.5 | 37 50 | 2.75 | 2.92 | 1.15 | 11 | | | | |
| 1901+050 | 19 01 16 | 05 05 19 | 0.32 | 1.46 | 1160 | 01 20 | 05 10 | 1.40 | 83.76 | 1.01 | 16 | | | | [3] |
| 1843-032 | 18 43 21 | -03 27 04 | 0.32 | 1.36 | 215 | 43 22 | 26 50 | 0.50 | 0.63 | 0.34 | 11 | | | | |
| 2314+595 | 23 14 37 | 59 54 23 | 0.32 | 1.26 | 670 | 14 38 | 54 20 | 1.50 | 94.59 | | 17 | | | | [3] |
| 1902+053 | 19 02 22 | 05 38 05 | 0.32 | 1.29 | 203 | 02 23.5 | 38 20 | 2.20 | | 0.45 | 18 | | | | [3] |
| 1906+053 | 19 06 15 | 05 31 10 | 0.31 | 1.28 | 964 | 06 15 | 31 15 | 1.40 | 93.70 | | 4 | | | | |
| Sum | | | | | 23234 | | | 131.70 | 1227.08 | 23.90 | | | | | |
| No.IRAS Sources | | | | | 31 | | | 31 | 31 | 31 | | | | | |
| Avg. per IRAS Source | | | | | 750 | | | 4.25 | 39.58 | 0.77 | | | | | |
| Tot. no. of Sources | | | | | 39 | | | 39 | 39 | 39 | | | | | |
| Avg. per Source | | | | | 596 | | | 3.38 | 31.46 | 0.61 | | | | | |
| (b) SOURCES WITH OFFSET > 3′. | | | | | | | | | | | | | | | |
| 1907+083 | 19 07 46 | 08 39 14 | 0.50 | 1.15 | 700 | | | 7.12 | 10.83 | 5.00 | 1 | | | | [3] |

TABLE 2—*Continued*

| IRAS Number | R.A. h.m.s. | Dec d.m.s. | Log 25/12 | Log 60/25 | 100μm Jy | R.A. Radio | Dec Radio | Peak mJy | Total mJy | Offset arcmin | Com | CH$_3$OH 6.6 | OH 1.6 | OH 4.7 | H$_2$O 22GHz |
|---|---|---|---|---|---|---|---|---|---|---|---|---|---|---|---|
| (c) SOURCES REJECTED |
| 1757-223 | 17 57 54 | -22 36 17 | 0.47 | 1.24 | 357 | | | 1.60 | | | 19 | | | | |
| 1854+011 | 18 54 27 | 01 12 19 | 0.46 | 1.46 | 987 | | | | | | 19 | | | | [3] |
| 1813-134 | 18 13 38 | -13 47 39 | 0.43 | 1.25 | 349 | | | 0.50 | | | 19 | | | | [3] |
| 1855+004 | 18 55 51 | 00 41 28 | 0.41 | 1.09 | 260 | | | 0.50 | | | 4 | | | | |
| 1806-202 | 18 06 06 | -20 25 42 | 0.36 | 1.24 | 27400 | G10.6 | | 12.00 | | | 4 | | | | |
| 1850+002 | 18 50 27 | 00 25 05 | 0.36 | 1.22 | 304 | | | | | | 19 | | | | |
| 1825-122 | 18 25 45 | -12 26 31 | 0.35 | 1.06 | 203 | | | 1.40 | | | 19 | | | | |
| 0012+605 | 00 12 48 | 60 58 01 | 0.35 | 1.01 | 405 | | | 75.00 | | | 20 | | | | |
| 1812-120 | 18 12 19 | -12 03 42 | 0.32 | 1.39 | 461 | | | 0.60 | | | 19 | | | | [3] |
| | | Average | 0.39 | 1.22 | 3414 | | | | | | | | | | |

COMMENTS
1 Complex                    2 Cometary shape             3 Diameter 15"
4 Very diffuse               5 Cometary, D=30"            6 Diameter 5"
7 Diameter 7"                8 Very weako                 9 Cometary,15"x20"
10 Other sources nearby      11 Not resolved              12 Diameter,20"x30"
13 Diameter,40"              14 Barely resolved           15 Diameter,7"
16 Diameter,45"              17 Diameter,1'               18 Diameter,20"
19 Not detected              20 Very complex

REFERENCES.— (3) Palla, et al. 1991; (4) Henning, et al. 1992

TABLE 3

Spectra of Objects Associated with IRAS Point Sources.

| Log(Average Values) | | | | Derived Flux Densities | | | | Av. Flux |
|---|---|---|---|---|---|---|---|---|
| $F_{25}/F_{12}$ | $F_{60}/F_{25}$ | $F_{100}/F_{60}$ | $F_{60}/F_{12}$ | $12\mu m$ | $25\mu m$ | $60\mu m$ | $100\mu m$ | $100\mu m$ |
| HII Regions | | Number: | 28 | | | | | |
| 0.61(0.32) | 0.97(0.40) | 0.31(0.38) | 1.58(0.37) | 1.30 | 5.36 | 49.46 | 100.00 | 1026 |
| Reflection Nebulae | | Number: | 23 | | | | | |
| 0.02(0.28) | 0.99(0.49) | 0.62(0.34) | 0.99(0.49) | 2.33 | 2.45 | 24.13 | 100.00 | 126.70 |
| Planetary Nebulae | | Number: | 36 | | | | | |
| 0.89(0.16) | 0.02(0.32) | -0.04(0.35) | 0.19(0.31) | 12.25 | 95.01 | 100.00 | 91.55 | 15.15 |

Note.—The numbers contained in parentheses are Standard Deviations